\documentclass[%
 reprint,
superscriptaddress,
 amsmath,amssymb,
 aps,
prb,
]{revtex4-2}

\usepackage[colorlinks=true,linkcolor=blue,urlcolor=blue,citecolor=blue]{hyperref}
\usepackage{graphicx}
\usepackage{dcolumn}
\usepackage{bm}
\usepackage{multirow}
\usepackage{float}
\usepackage{setspace}

\begin{document}


\title{Orbital doublet driven even-spin Chern insulators}


\author{Lu Liu}
\thanks{Lu Liu and Yuntian Liu contributed equally to this work.}
\affiliation{Department of Physics and Shenzhen Institute for Quantum Science and Engineering (SIQSE), Southern University of Science and Technology, Shenzhen 518055, China}
 \affiliation{Laboratory for Computational Physical Sciences (MOE),
 State Key Laboratory of Surface Physics, and Department of Physics,
  Fudan University, Shanghai 200433, China}

\author{Yuntian Liu}
\thanks{Lu Liu and Yuntian Liu contributed equally to this work.}
\affiliation{Department of Physics and Shenzhen Institute for Quantum Science and Engineering (SIQSE), Southern University of Science and Technology, Shenzhen 518055, China}

\author{Jiayu Li}
\affiliation{Department of Physics and Shenzhen Institute for Quantum Science and Engineering (SIQSE), Southern University of Science and Technology, Shenzhen 518055, China}

\author{Hua Wu}
\email{Corresponding author. wuh@fudan.edu.cn}
\affiliation{Laboratory for Computational Physical Sciences (MOE),
 State Key Laboratory of Surface Physics, and Department of Physics,
 Fudan University, Shanghai 200433, China}
 \affiliation{Shanghai Qi Zhi Institute, Shanghai 200232, China}
\affiliation{Collaborative Innovation Center of Advanced Microstructures,
 Nanjing 210093, China}

\author{Qihang Liu}
\email{Corresponding author. liuqh@sustech.edu.cn}
\affiliation{Department of Physics and Shenzhen Institute for Quantum Science and Engineering (SIQSE), Southern University of Science and Technology, Shenzhen 518055, China}
\affiliation{Guangdong Provincial Key Laboratory of Computational Science and Material Design, Southern University of Science and Technology, Shenzhen 518055, China}

\date{\today}

\begin{abstract}
Quantum spin Hall insulators hosting edge spin currents hold great potential for low-power spintronic devices. In this paper, we present a universal approach to achieve a high and near-quantized spin Hall conductance plateau within a sizable bulk gap. Using a nonmagnetic four-band model Hamiltonian, we demonstrate that an even-spin Chern (ESC) insulator can be accessed by tuning the sign of spin-orbit coupling (SOC) within a crystal symmetry-enforced orbital doublet. With the assistance of a high spin Chern number of $C_{S}=-2$ and spin $U$(1) quasisymmetry, this orbital-doublet-driven ESC phase is endowed with the near-double-quantized spin Hall conductance. We identify 12 crystallographic point groups supporting such a sign-tunable SOC. Furthermore, we apply our theory to realistic examples, and show the phase transition from a trivial insulator governed by positive SOC in the RuI$_{3}$ monolayer to an ESC insulator dominated by negative SOC in the RuBr$_{3}$ monolayer. This orbital-doublet-driven ESC insulator, RuBr$_{3}$, showcases nontrivial characteristics including helical edge states, near-double-quantized spin Hall conductance, and robust corner states. Our work provides different pathways in the pursuit of the long-sought quantum spin Hall insulators.
\end{abstract}

\maketitle
 
\section*{Introduction}
Two-dimensional (2D) quantum spin Hall (QSH) insulators have garnered significant interest for their promising applications in spintronics and magnetoelectronics \cite{QSH_gra,ZSC_Science2006,TI_RMP2010,TI_RMP2011}. They manifest topologically protected helical edge states where the spin is locked to the momentum through spin-orbit coupling (SOC) and time-reversal symmetry (TRS), providing dissipationless spin transports ideal for low-power magnetic memory devices. The first predictions of realistic QSH insulators identified graphene \cite{QSH_gra} and the HgTe quantum well \cite{ZSC_Science2006} as candidates, each characterized by a SOC-induced inverted bulk gap along with a pair of helical edge states within this gap. This topological phase is generally characterized by the topological invariant $Z_{2}=1$, which also serves as the symmetry indicator for TRS-preserved systems \cite{Z2_2005}. Over the years, this $Z_{2}=1$ topological phase has been observed in several quantum wells \cite{HgTe_exp1,HgTe_exp2,InAs_exp} and pristine 2D materials such as WTe$_{2}$, bismuthene, Na$_{3}$Bi, and germanene \cite{WTe2_NP1,WTe2_NP2,WTe2_science, bismuthene, Na3Bi, germanene}.

In addition to the $Z_{2}$ index, the spin Chern number $C_{S}$, also established as a topological invariant, is directly related to the number of pairs of helical edge states \cite{Haldane2006}. In particular, when the real-spin component $S_{z}$ remains preserved, $C_{S}$ defines the quantized spin Hall conductance (SHC) as $\sigma_{xy}^{S}=C_{S}\frac{e}{2\pi}$. These two invariants are related by $Z_{2}=$ mod ($C_{S},2$). Therefore, QSH insulators with two pairs of helical edge states in the $|C_{S}|=2$ regime are considered to be trivial within conventional $Z_{2}$ classification. However, experiments have observed near-double-quantized conductance in twisted bilayers WSe$_{2}$ and MoTe$_{2}$ \cite{twist_arxiv1, twist_arxiv2}, demonstrating that QSH effects can indeed manifest in even-spin Chern (ESC) insulators. Regarding the absence of spin $U$(1) symmetry in realistic materials and the consequent lack of exact quantization of SHC \cite{Z2_2005, Haldane2006, wenxg}, we have recently emphasized the pivotal role of spin $U$(1) quasisymmetry for the near-quantization of SHC in TRS-preserved $Z_{2}=1$ or such $Z_{2}=0$ systems, as well as TRS-broken cases \cite{liu_quasi}. Beyond theoretical predictions of magnetic high spin Chern insulators \cite{ezawa_high_2013}, and the ESC phase in monolayer $\alpha$-Sb/Bi \cite{esc_prb, baokai_arxiv1, Sb2023} and magnetic Fe$_{2}$BrMgP monolayer and TiTe bilayer \cite{yzq}, we have predicted near-double-quantized SHC in twisted bilayer transition metal dichalcogenides and monolayer RuBr$_{3}$ \cite{liu_quasi} protected by spin $U$(1) quasisymmetry. In this paper, we will present a general approach to realize an ESC phase with a symmetry-protected near-double-quantized SHC within a large bulk gap, which would be an ideal platform for observing QSH effects and further promote applications of QSH insulators.

\begin{figure}[t]
  \centering
\includegraphics[width=8.2cm]{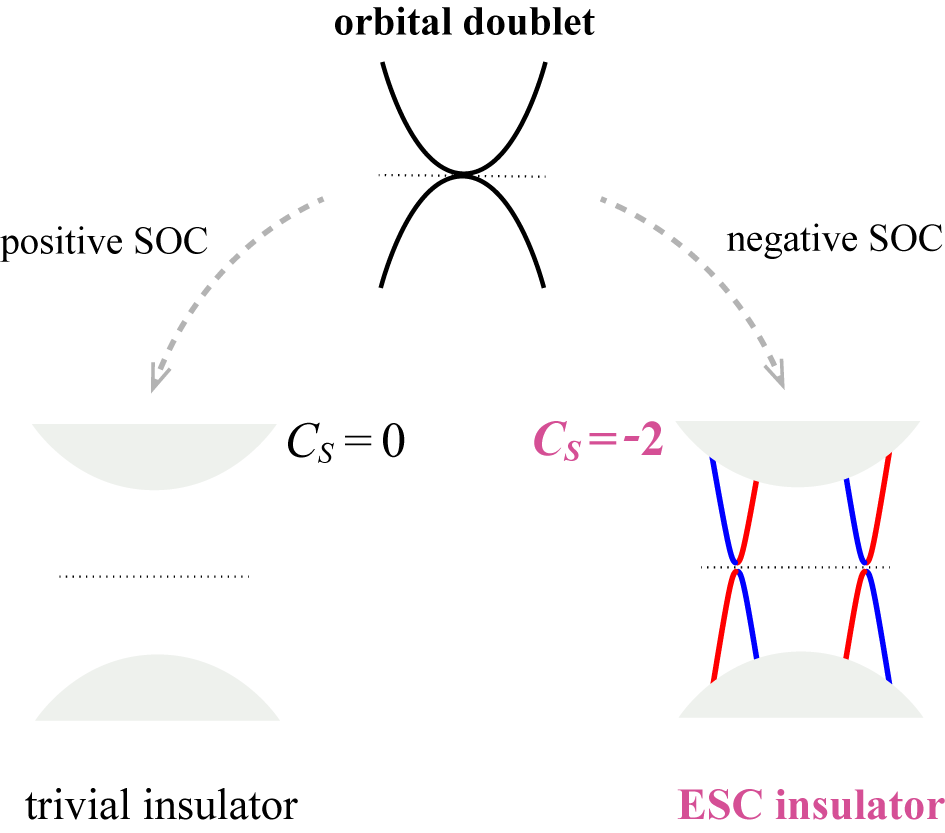}
  \caption{Schematic for designing ESC insulators with $C_{S}=-2$: by tuning the sign of SOC within an orbital doublet from positive to negative, a phase transition from $C_{S}=0$ to $C_{S}=-2$ can be realized.}
  \label{fig:1}
\end{figure}

First, using a nonmagnetic four-band model Hamiltonian, we demonstrate that an ESC phase with $C_{S}=-2$ can be accessed by tuning the sign of SOC within a crystal symmetry-enforced orbital doublet, as shown in Fig. \ref{fig:1}. Such an orbital-doublet-driven ESC phase is endowed with two notable features: (i) a sizable bulk gap opened by first-order spin-preserved SOC; (ii) a high near-quantized SHC approaching $-$2 (in unit of $e/2\pi$) protected by spin $U$(1) quasisymmetry. Thereafter, we enumerate 12 crystallographic point groups supporting the orbital doublets with sign-alterable SOC effects. Furthermore, we present realistic examples to demonstrate our theory. As shown below, a trivial insulator driven by positive SOC transforms into a nontrivial ESC insulator induced by negative SOC, as observed in the transition from monolayer RuI$_{3}$ to RuBr$_{3}$. In addition to the topologically nontrivial features such as the near-double-quantized SHC and two pairs of helical edge states, we further show robust in-gap corner states that is associated with the slightly gapped edge states in RuBr$_{3}$.

\section*{Computational Details}
Density functional theory calculations are performed using the full-potential augmented plane wave plus the local orbital code (Wien2k) \cite{wien2k}. The optimized lattice constants of RuI$_{3}$, RuBr$_{3}$, RuCl$_{3}$, and RuF$_{3}$ monolayers are $a=b=$ 6.667, 6.159, 5.747, and 4.827 \AA, respectively. A vacuum slab of 15 \AA~is set along the $c$ axis for both systems. The muffin-tin sphere radii are chosen to be 2.2, 2.4, 2.1, and 1.5 bohrs for Ru, I/Br, Cl, and F atoms, respectively. The cutoff energy of 14 Ry is set for plane wave expansions of interstitial wave functions. We use the 11$\times$11$\times$1 $k$-mesh for integration over the Brillouin zone. SOC is included by the second variational method with scalar relativistic wave functions. Electron correlation of Ru $4d$ electrons is taken into account by adopting a typical Hubbard $U$ of 2 eV and a Hund's exchange of 0.5 eV \cite{Anisimov_1993}. The Wannier functions of Ru 4$d$, I 5$p$, and Br 4$p$ orbitals are constructed using Wien2wannier \cite{wien2wannier} and WANNIER90 \cite{wannier90} without performing maximally localized procedures. The topological edge states and SHC are calculated by the iterative Green’s function and the Kubo formula \cite{SHC_cal}, respectively, as implemented in WannierTools package \cite{wt}. Since the RuCl$_{3}$ monolayer exhibits electronic structures and topological characters similar to those of RuBr$_{3}$, and the metallic RuF$_{3}$ monolayer is beyond our interests, we present the results for RuI$_{3}$ and RuBr$_{3}$ monolayers in the main text, and those for RuCl$_{3}$ and RuF$_{3}$ monolayers in Fig. \ref{fig:7} in the Appendix B.

\section*{Results}
\subsection*{I. Symmetry and model of even spin Chern phase}

\begin{figure}[t]
  \centering
\includegraphics[width=8.5cm]{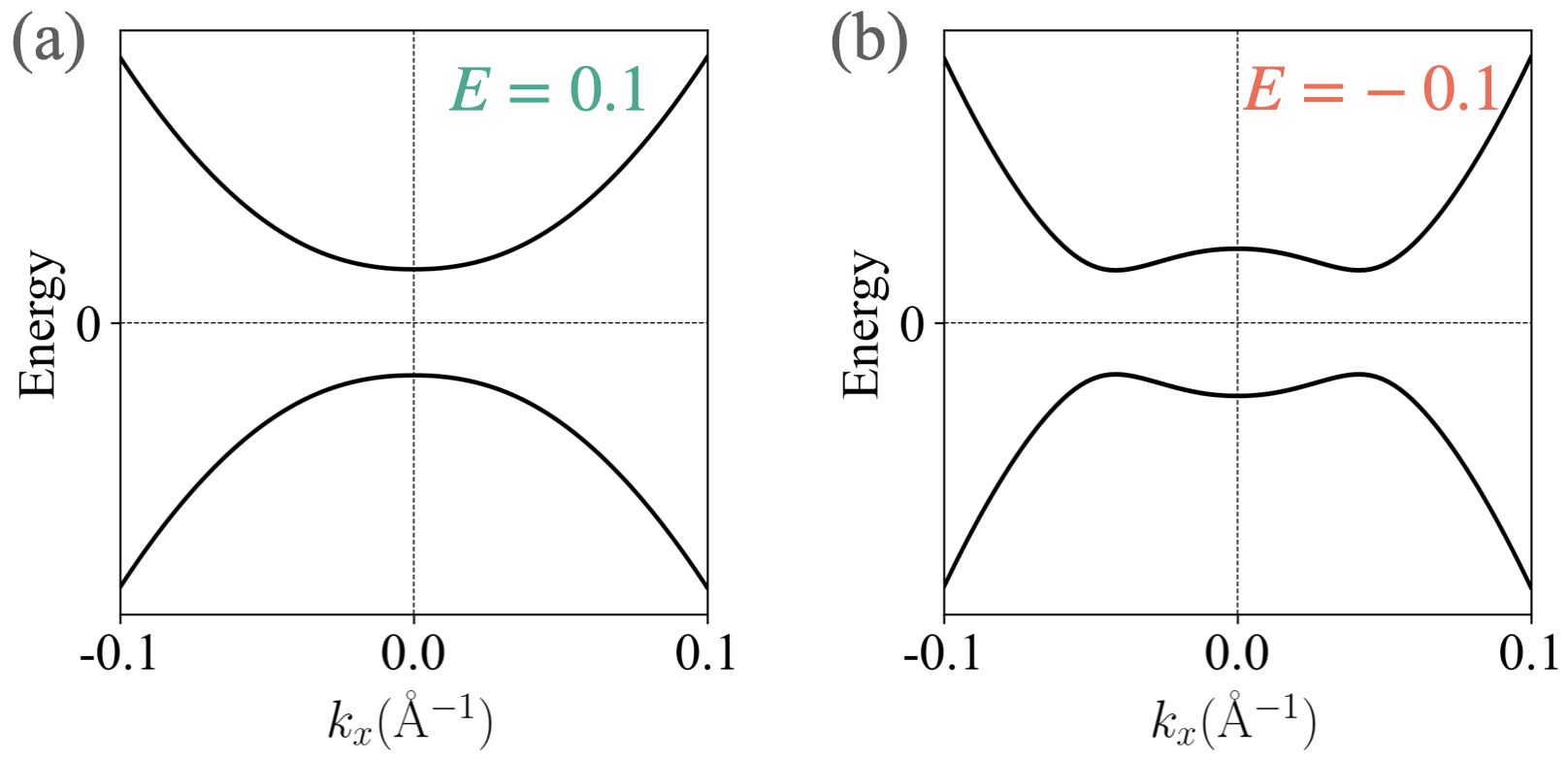}
  \caption{Band structures of the model Hamiltonian in Eq. (\ref{equation1}) with the parameters $A=B=0, C=D=0.3$: from (a) $E=0.1$ to (b) $E=-0.1$, the band inversion marks a phase transition from a trivial insulator to an ESC insulator with $C_{S}=-2$.}
  \label{fig:2}
\end{figure}

To begin with, we will show that a nontrivial ESC phase can be realized within a nonmagnetic four-band model Hamiltonian based on an orbital doublet that is characterized by a 2D irreducible representation (irrep). We consider a typical doublet formed by $p_{x}$ and $p_{y}$ orbitals as $p_{\pm}= (p_{x}\pm i p_{y})/\sqrt{2}$, where the subscript $+$/$-$ denotes orbital angular momentum $l_{z}=+1$/$-1$. To generate a 2D irrep furnished by the $p_{\pm}$ doublet, here we consider a $D_{6h}$ point group, of which the generators are three-fold rotation symmetry $C_{3z}$ along the $z$ axis, two-fold rotation symmetry $C_{2z}$/$C_{2y}$ along the $z$/$y$ axis, and space inversion symmetry $I$. In the basis of $\{|p_{+},\uparrow\rangle, |p_{-},\uparrow\rangle, |p_{+},\downarrow\rangle, |p_{-},\downarrow\rangle\}$, the representation of symmetry operations is given by $C_{3z}=e^{i\frac{\pi}{3}\sigma_{z}}\otimes e^{i\frac{2\pi}{3}\tau_{z}}$, $C_{2z}=e^{i\frac{\pi}{2}\sigma_{z}}\otimes -\tau_{0}$, $C_{2y}=e^{i\frac{\pi}{2}\sigma_{y}}\otimes-\tau_{x}$, $I=\mathbb{I}_{2 \times 2}\otimes -\mathbb{I}_{2 \times 2}$, and TRS $T=\mathcal{K}\cdot i \sigma_{y}  \otimes \tau_{x}$, where $\mathcal{K}$ is the complex conjugation operator, $\mathbb{I}_{2 \times 2}$ is a $2 \times 2$ identity matrix, and $\sigma_{x,y,z}$ and $\tau_{x,y,z}$ are Pauli matrices for spin and orbital degrees of freedom, respectively. By imposing those symmetries, we derive the generic form of the effective Hamiltonian as follows:
\begin{equation}
\label{equation1}
\begin{aligned}
H(\bm{k})=&\epsilon_{0}(\bm{k}) \mathbb{I}_{4 \times 4} + C[(k_{x}^{2}-k_{y}^{2})\sigma_{0}\otimes\tau_{x}+2k_{x}k_{y}\sigma_{0}\otimes\tau_{y}]\\
& + D (k_{x}^{2}+k_{y}^{2})\sigma_{z}\otimes\tau_{z}+E \sigma_{z}\otimes\tau_{z}
\end{aligned}
\end{equation}
with $\epsilon_{0}(\bm{k})=A-B(k_{x}^{2}+k_{y}^{2})$. Note that the symmetry preserves the term $E \sigma_{z}\otimes\tau_{z}$ which is contributed by the first-order spin-preserved SOC. The resulting electronic structure consists of two sets of doubly degenerate bands protected by $I$ and $T$ symmetry, yielding an energy gap 2$E$. This bulk gap 2$E$, primarily opened by first-order SOC, can reach $\sim$100 meV to against thermal fluctuation and local disorder. Furthermore, the change of the sign of $E$ from positive to negative marks a phase transition accompanied by band inversion, as shown in Fig. \ref{fig:2}. Note that such band inversion does not change the $Z_{2}$ index of the system because the wavefunctions of the lowest conduction band and the highest valence band at $\Gamma$ share the same parity. However, we find that such a band inversion signifies a topological phase transition from a trivial insulator to a nontrivial ESC insulator characterized by $C_{S}=-2$ (see Appendix A).

We note that such a topological phase transition, driven by altering the sign of the SOC within orbital doublets, can be achieved by orbital engineering. Specifically, some $d$-orbital doublets, undergoing transformations identical to the $p_{\pm}$ doublet but with an opposite $l_{z}$, can contribute negative SOC in contrast to the positive one within $p_{\pm}$. We identify 12 crystallographic point groups that can support the sign-alterable SOC within specific orbital doublets, as listed in Table \ref{tb1}. For instance, the orbital doublet $d_{\pm2}=|l_{z}=\pm2\rangle$ is supported by $(C_{3h}, D_{3h}, C_{6}, C_{6v}, C_{6h}, D_{6}, D_{6h})$ point groups. Under the rotational symmetries that can distinguish the two states in an orbital doublet, the $d_{-2}$ state transform as $p_{+}$, and $d_{+2}$ transforms as $p_{-}$, e.g., the symmetry operation $C_{3z}$ introduces a phase factor $e^{-i\frac{2\pi}{3}\tau_{z}}$ to $d_{\pm2}$ but an opposite phase factor $e^{i\frac{2\pi}{3}\tau_{z}}$ to $p_{\pm}$. Therefore, the $d_{\pm2}$ doublet will yield a negative splitting when SOC emerges, in contrast to the positive SOC-splitting in $p_{\pm}$. Similarly, the $e_{\pm}^{\prime}$ doublet supported by ($C_{3}, C_{3v}, D_{3}, D_{3d}, S_{6}$) point groups is formulated as $e^{\prime}_{\pm}=\pm \cos\alpha|l_{z}=\pm2\rangle-\sin\alpha|l_{z}=\mp1\rangle$, where $\sin^{2}\alpha$ varies from 0 to 1/3 depending on local $d$-orbital environments \cite{Whangbo}. The $e_{\pm}^{\prime}$ transforms as $p_{\mp}$, and thus also provides a negative SOC. Note that two other $d$-orbital doublets listed in Table \ref{tb1}, specifically $d_{\pm}=|l_{z}=\pm1\rangle$ and $e^{\prime}_{g\pm}=\sin\alpha|l_{z}=\mp2\rangle \mp\cos\alpha|l_{z}=\pm1\rangle$, both yield the positive SOC just as that in $p_{\pm}$.

\begin{table}[b]
\renewcommand{\arraystretch}{1.2}
 \caption{Crystallographic point groups that permit orbital doublets with a sign-tunable SOC (both positive and negative).}
  \label{tb1}
  \begin{tabular}{p{4.9cm}r@{\extracolsep{0.7cm}}c}
  \hline\hline
point groups & doublets & SOC-sign \\ \hline
$C_{3h}, D_{3h}, C_{6}, C_{6v}, C_{6h}, D_{6}, D_{6h}$ & $p_{\pm},d_{\pm}$  & + \\
& $d_{\pm2}$ &  $-$ \\\hline
$C_{3}, C_{3v}, D_{3}, D_{3d}, S_{6}$  & $p_{\pm},e_{g\pm}^{\prime}$ & + \\ 
    & $e_{\pm}^{\prime}$ &  $-$   \\ \hline
 \end{tabular}
\end{table}

We emphasize that among the 12 crystallographic point groups in Table \ref{tb1}, while lowering symmetries from the highest symmetric point group $D_{6h}$ [Eq. (\ref{equation1})] may introduce additional terms, the low-energy physics at the $\Gamma$ point remains intact. For instance, in point group $D_{3d}$, the term $F[(k_{x}^{2}-k_{y}^{2})\sigma_{x}\otimes\tau_{z}+2k_{x}k_{y}\sigma_{y}\otimes\tau_{z}]$ emerges \cite{liu_quasi}, serving as spin-mixing perturbations. More notably, within the eigenspace of the model Hamiltonian in Eq. (\ref{equation1}), which is spanned by an orbital doublet combined with electron spin, spin $U$(1) quasisymmetry is present to eliminate the first-order spin-mixing perturbation \cite{liu_quasi, quasi_JY}. Such a symmetry plays a pivotal role for protecting QSH effects in realistic materials. Consequently, despite a trivial $Z_{2}$ index, the ESC systems described by our model can exhibit a near-double-quantized SHC plateau within a sizable bulk gap. In addition, the edge state would open a small gap by spin-mixing perturbation. These features are further confirmed by realistic 2D examples presented in the following section.

\subsection*{II. Realistic materials with tunable SOC}

\begin{figure}[t]
  \centering
\includegraphics[width=8.5cm]{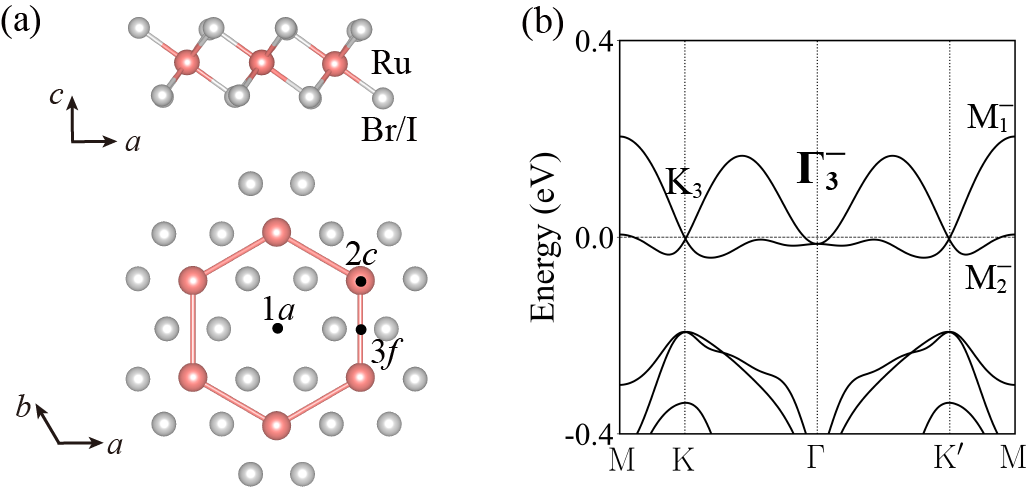}
  \caption{(a) Crystal structure of RuI$_{3}$ and RuBr$_{3}$ monolayers with Ru and I/Br atoms represented by red and gray balls, respectively. In the bottom panel, the 1$a$, 2$c$, and 3$f$ maximal Wyckoff positions within the $c=0$ plane are denoted. (b) Band structures of RuI$_{3}$ monolayer without SOC. The Fermi level is set at the zero energy.}
  \label{fig:3}
\end{figure}

We take the RuI$_{3}$ and RuBr$_{3}$ monolayers as examples to demonstrate an ESC phase that is accessible through tuning the sign of SOC. The three-dimensional form of RuI$_{3}$ has been crystallized in a rhombohedral structure with space group $R\bar{3}$ \cite{Cava_2022am}, and its 2D counterpart is in the space group $P\bar{3}1m$, providing the little point group $D_{3d}$ at the $\Gamma$ point, which is included in Table \ref{tb1}. Recent studies have shown that, due to intricate SOC effects combined with strong Ru-I hybridization, RuI$_{3}$ exhibits paramagnetic behavior and undergoes a metal-to-insulator transition from bulk to monolayer \cite{Cava_2022am,Nawa_2021JPSJ,Kaib_2022arx,Zhang_2022PRB,lliu_PRB}. Therefore, RuI$_{3}$ monolayer would be a great platform for investigating SOC effects on topological characteristics. Moreover, RuBr$_{3}$ monolayer is also of interest for the variation of the relative importance of the SOC at the Ru and ligand Br/I sites, and for the possibly new topological properties.

\begin{figure}[t]
  \centering
\includegraphics[width=8.5cm]{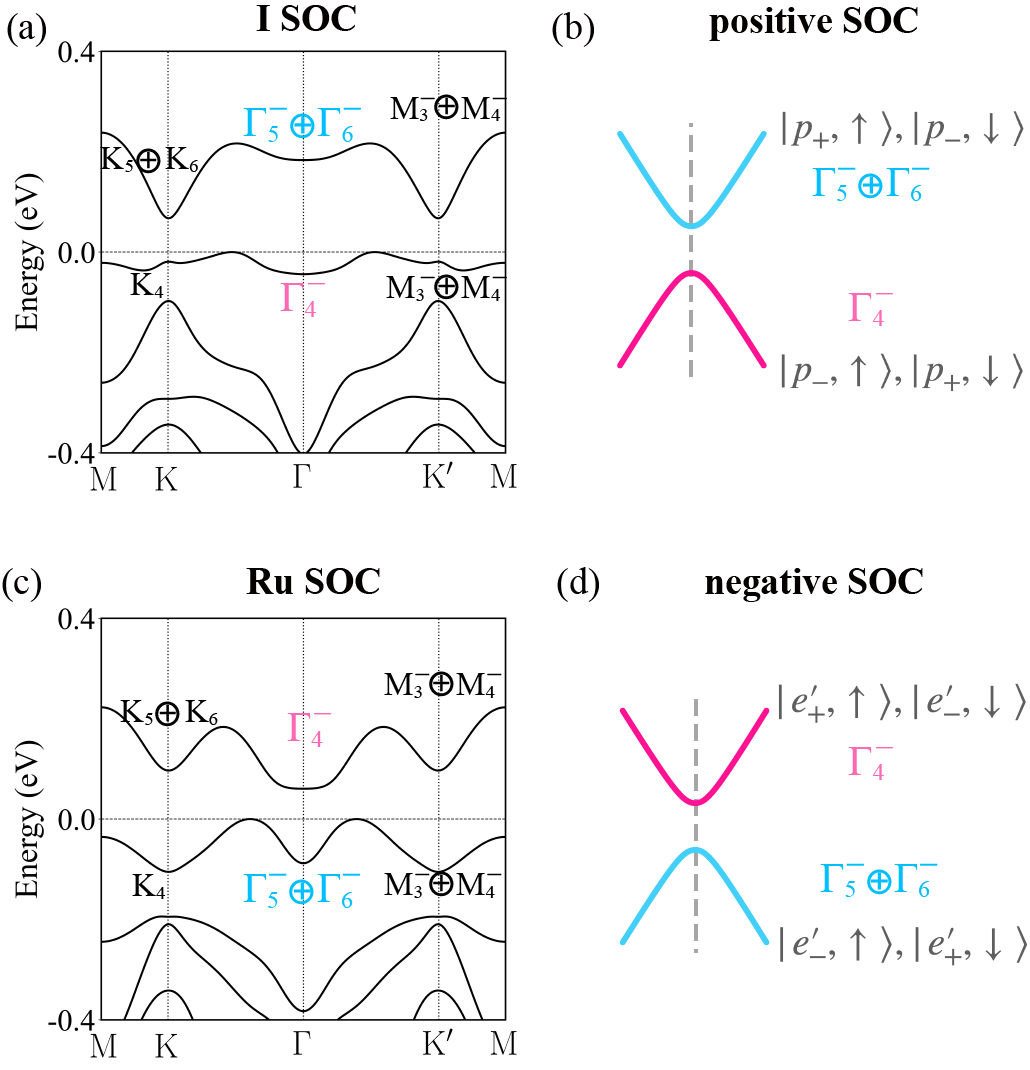}
  \caption{Band structures of RuI$_{3}$ monolayer with (a) only I SOC active and (c) only Ru SOC active. Combined with electron spin, the orbital doublets (b) $p_{\pm}$ of I $5p$ states and (d) $e^{\prime}_{\pm}$ of Ru $4d$-$t_{2g}$ states form the co-irreps around the Fermi level at the $\Gamma$ point, undergoing positive and negative SOC-splitting, respectively.}
  \label{fig:4}
\end{figure}

\begin{figure}[t]
  \centering
\includegraphics[width=8.5cm]{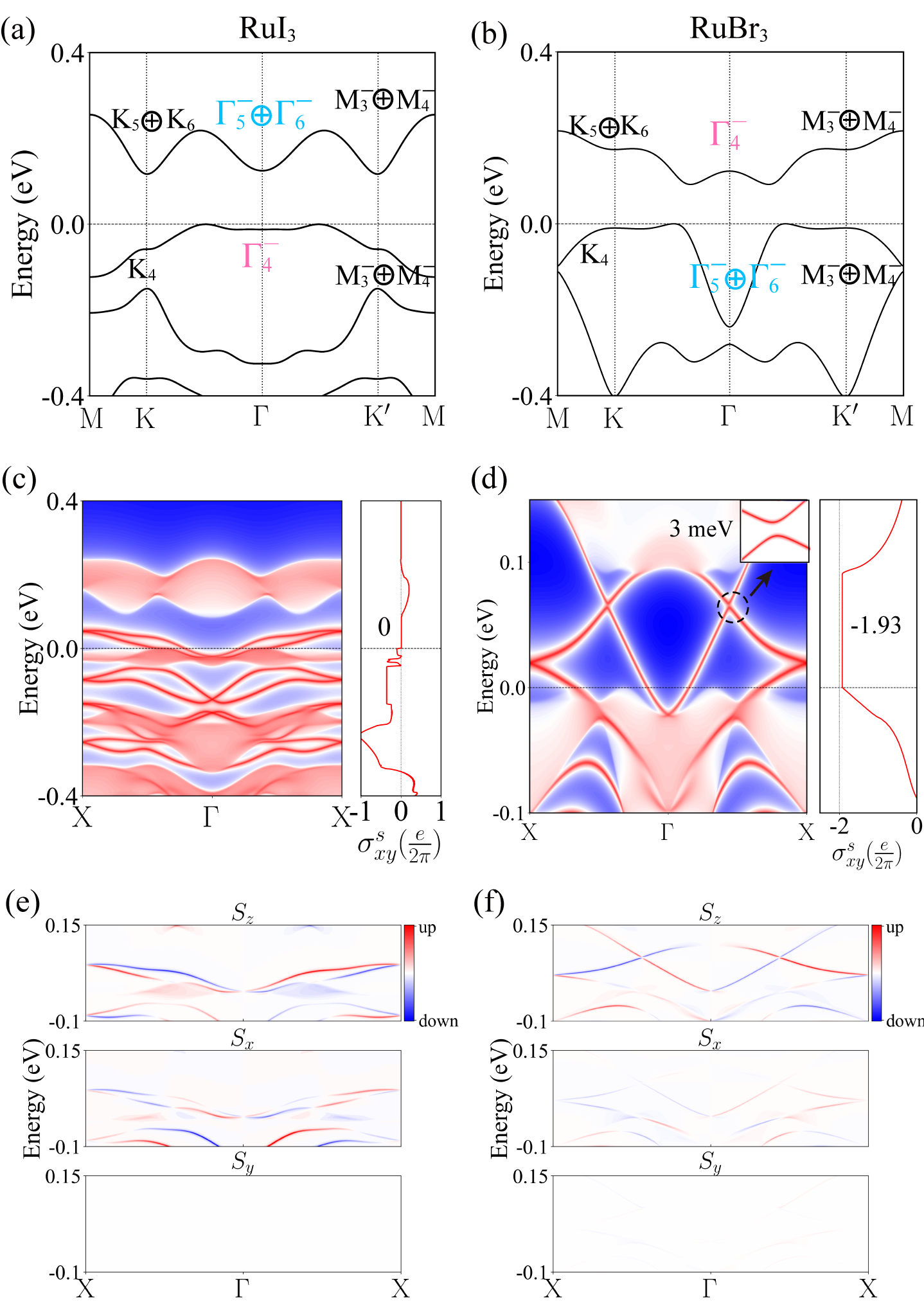}
  \caption{(a) Band structures with SOC, (c) edge spectrum and SHC, and (e) spin components of the edge states for RuI$_{3}$ monolayer; (b), (d), and (f) corresponding results for RuBr$_{3}$ monolayer.}
  \label{fig:5}
\end{figure}

We first present the band structures of RuI$_{3}$ monolayer in the absence of SOC. Figure \ref{fig:3}(b) illustrates that, without SOC, two isolated bands around the Fermi level form crossings at $\Gamma$ and K points. This band degeneracy is protected by the crystal symmetry and can be lifted by SOC. As shown in Figs. \ref{fig:4} and \ref{fig:5}(a), a bulk gap is opened when SOC emerges, signifying the RuI$_{3}$ monolayer as a band insulator [individual I and Ru SOC effect in Figs. \ref{fig:4}(a) and \ref{fig:4}(c), and joint one in Fig. \ref{fig:5}(a)]. To characterize the topological phase of RuI$_{3}$ monolayer, we calculate the $Z_{2}$ index by computing the parity eigenvalues of valence bands at two time-reversal-invariant momenta \cite{FuKane_Z2}, namely $\Gamma$ and M. The same parity at $\Gamma$ and M yields a $Z_{2}=0$. As a result, we find that RuI$_{3}$ monolayer is a $Z_{2}$ trivial insulator.

It is worth to note that within the category of topologically trivial insulators, there exists a special subgroup known as obstructed atomic insulators (OAIs), as proposed based on topological quantum chemistry (TQC) theory \cite{TQC_nature,TQC_prb,OAI_arxiv,OAI_scibu,lyt_OAI}. Within the TQC framework, for topologically trivial insulators, the band representation (BR) of all occupied bands is a sum of elementary band representations (EBRs) induced from atomic orbitals at maximal Wyckoff positions. And OAI refers to the situation that some of those Wyckoff positions are empty sites without atoms occupied. By calculating the BR of valence bands, we find that the BR decompositions of RuI$_{3}$ monolayer have to include an EBR at empty Wyckoff position $1a$, i.e., the center of the honeycomb lattice, see Fig. \ref{fig:3}(a). Therefore, RuI$_{3}$ monolayer falls into the category of OAIs. This is also captured by the emergence of obstructed metallic edge states, as shown in Fig. \ref{fig:5}(c), which appears when one cuts the edge containing the obstructed $1a$ site. 

We now take a close look at SOC effects. As shown above, SOC-splitting is responsible for the band gap of RuI$_{3}$ monolayer. When we examine the individual contributions of SOC from Ru and I elements, we find that the band splitting in RuI$_{3}$ is primarily driven by I SOC, as evidenced by the same co-irrep feature of the lowest conduction bands and highest valence bands, i.e., ($\Gamma_{5}^{-}\oplus \Gamma_{6}^{-}$)-over-$\Gamma_{4}^{-}$, for both Figs. \ref{fig:4}(a) and \ref{fig:5}(a). In contrast, when Ru SOC is considered independently, as shown in Fig. \ref{fig:4}(c), the band gap at $\Gamma$ point is inverted, yielding a negative splitting $\Gamma_{4}^{-}$-over-($\Gamma_{5}^{-}\oplus \Gamma_{6}^{-}$). Such SOC-sign-change behavior is well predicted as the case of $D_{3d}$ in Table \ref{tb1}.

Despite the fact that either positive or negative SOC-splitting of the orbital doublet does not change $Z_{2}$, the model Hamiltonian in Eq. (\ref{equation1}) predicts that the SOC-sign-change triggers a topological phase transition between the trivial $C_{S}=0$ phase and the nontrivial $C_{S}=-2$ phase. To provide a realistic material candidate for the latter case, we naturally move to RuBr$_{3}$, taking into account the reduced SOC strength associated with Br $4p$ electrons and their weaker hybridization with Ru $4d$ states as compared to I $5p$ electrons. As anticipated, our results show a band inversion from RuI$_{3}$ to RuBr$_{3}$, as evidenced by the SOC-induced splitting at $\Gamma$ shifting from a positive ($\Gamma_{5}^{-}\oplus \Gamma_{6}^{-}$)-over-$\Gamma_{4}^{-}$ configuration to a negative $\Gamma_{4}^{-}$-over-($\Gamma_{5}^{-}\oplus \Gamma_{6}^{-}$) one, see Figs. \ref{fig:5}(a) and \ref{fig:5}(b).

\subsection*{III. Nontrivial features in ESC insulator}

\begin{figure}[t]
  \centering
\includegraphics[width=8.5cm]{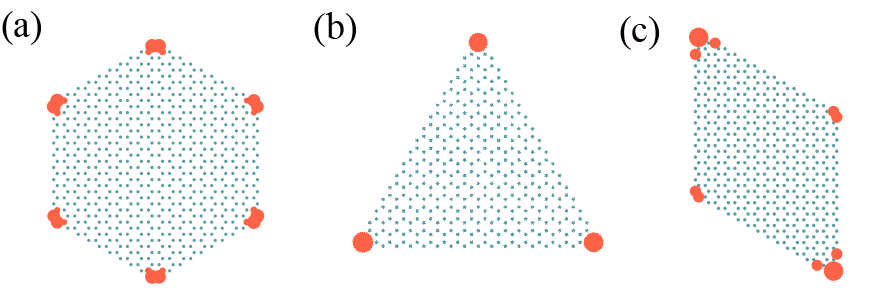}
  \caption{Spatial distributions of the state at the Fermi level for (a) hexagonal, (b) triangular, and (c) rhomboid-shaped nanodisks of RuBr$_{3}$.}
  \label{fig:6}
\end{figure}

Despite both RuI$_{3}$ and RuBr$_{3}$ belonging to the $Z_{2}=0$ phase, their distinct topological features are evident in the edge and SHC behaviors. In stark contrast to RuI$_{3}$, RuBr$_{3}$ exhibits four metallic edge states and two Dirac-like edge crossings, see Figs. \ref{fig:5}(c-f). A closer examination of these edge crossings reveals a small gap of 3 meV, which is opened by spin-mixing perturbations. Moreover, unlike the absent SHC in RuI$_{3}$, RuBr${_3}$ exhibits a SHC plateau within a large bulk gap of 130 meV, and the SHC value of $-$1.93 closely approaches the quantized value of $-$2. As discussed above, the transition from RuI$_{3}$ to RuBr${_3}$ involves a weakening of the positive SOC from the ligands $p$ orbitals, with the Ru negative SOC become dominant in RuBr${_3}$, giving rise to the topological phase transition from a trivial insulator with $C_{S}=0$ to an ESC insulator with $C_{S}=-2$. Note that these topologically nontrivial features in RuBr$_{3}$ are protected by a nonzero $C_{S}$ and spin $U$(1) quasisymmetry \cite{liu_quasi}. Thus, our findings highlight the orbital-doublet-driven ESC insulators, as described by our nonmagnetic four-band model Hamiltonian, as an ideal platform for realizing QSH effects.

In addition, we note that Ref. \cite{baokai_arxiv2} has predicted that in ESC insulators, general spin-mixing interactions that open the edge gap introduce a mass term on the edges, accompanied by a phase difference between the mass terms on adjacent edges. Such an edge mass-kink \cite{masskink} would give rise to corner localized charge and thus the in-gap corner modes, independent of the specific symmetry indicators or the geometry of nanodisks. Here, the ESC insulator RuBr$_{3}$ with a spin-mixing induced edge gap closely conforms to the case in Ref. \cite{baokai_arxiv2}. To check the existence of corner states, we construct nanodisks with hexagonal, triangular, and rhomboid shapes for RuBr$_{3}$ and plot the real-space distributions of the state at the Fermi level, which is determined by valence electron counting and resides within the energy range of the edge gap. As shown in Fig. \ref{fig:6}, we find that the in-gap states are well localized at the corners, independent of the geometry. Our results are in accordance with theoretical predictions about robust corner states in TRS-preserved $|C_{S}|=2$ systems \cite{baokai_arxiv2}. As a result, manifold nontrivial characteristics embedded in the orbital-doublet-driven ESC insulators, including helical edge states, high near-quantized SHC, and robust in-gap corner modes, enrich their potential applications spanning various fields.

\section*{Summary}
To summarize, we develop a nonmagnetic four-band model Hamiltonian based on a crystal symmetry-enforced orbital doublet. We propose a generic approach to realize a nontrivial ESC phase with $C_{S}=-2$ by tuning the sign of SOC within orbital doublets, which can be supported in 12 crystallographic point groups. Realistic 2D examples, specifically the evolution from RuI$_{3}$ monolayer to RuBr$_{3}$, demonstrate that a trivial $C_{S}=0$ insulator governed by positive SOC transforms into a nontrivial $C_{S}=-2$ insulator dominated by negative SOC. Moreover, we show that such orbital-doublet-driven ESC insulators manifest nontrivial features, including two pairs of helical edge states, high near-quantized SHC, and robust in-gap corner modes. Our work presents a universal strategy to design ESC insulators featuring a near-double-quantized SHC plateau within a large bulk gap, offering different insights into the exploration of QSH insulators.

\section*{Acknowledgments}
Q. L. acknowledges support by National Key R$\&$D Program of China under Grant No. 2020YFA0308900, National Natural Science Foundation of China under Grant No. 12274194, Guangdong Provincial Key Laboratory for Computational Science and Material Design under Grant No. 2019B030301001, Shenzhen Science and Technology Program under Grant No. RCJC20221008092722009, the Science, Technology and Innovation Commission of Shenzhen Municipality under Grant No. ZDSYS20190902092905285 and Center for Computational Science and Engineering of Southern University of Science and Technology. H. W. acknowledges support by the National Natural Science Foundation of China under Grants No. 12174062 and No. 12241402.

\newpage

\begin{widetext}
\appendix

\section{Spin Chern number for nonmagnetic four-band model}

The low-energy effective Hamiltonian under a $D_{6h}$ point group [Eq. (\ref{equation1})] does not include spin-mixing terms, so we could separate the spin-up and spin-down channels for simplicity. Then the Hamiltonian for the spin-up channel is as follows:
\begin{equation}
\label{equation2}
H(\bm{k})=\epsilon_{0}(\bm{k})\tau_{0}+ C[(k_{x}^{2}-k_{y}^{2})\tau_{x}+2k_{x}k_{y}\tau_{y}]+ \left[E+D (k_{x}^{2}+k_{y}^{2})\right]\tau_{z}
\end{equation}
where $\epsilon_{0}(\bm{k})=A-B(k_{x}^{2}+k_{y}^{2})$. By substituting $k_{x}=k\cos\phi$ and $k_{y}=k \sin\phi$, Eq. (\ref{equation2}) can be rewritten as
\begin{equation}
\label{equation3}
H(\bm{k})=\left(\begin{array}{cc}
\epsilon_{0}(\bm{k})+\cos\theta & \sin\theta e^{-i2\phi} \\
\sin\theta e^{i2\phi} & \epsilon_{0}(\bm{k})-\cos\theta
\end{array}\right)
\end{equation}
where $\cos\theta=	\frac{E+D k^{2}}{[C^{2}k^{4}+(E+Dk^{2})^2]^{\frac{1}{2}}}$ and $\sin\theta=\frac{Ck^{2}}{[C^{2}k^{4}+(E+Dk^{2})^2]^{\frac{1}{2}}}$. The wave functions are given by
\begin{equation}
\label{equation5}
\left|-\right\rangle=\left(\begin{array}{c}
e^{-i 2\phi} \sin \frac{\theta}{2} \\
-\cos \frac{\theta}{2}
\end{array}\right) \quad \text { and } \quad\left|+\right\rangle=\left(\begin{array}{c}
e^{-i 2\phi} \cos \frac{\theta}{2} \\
\sin \frac{\theta}{2}
\end{array}\right)
\end{equation}
with corresponding energies $E_{\pm}=\epsilon_{0}(\bm{k})\mp \sqrt{C^{2}k^{4}+(E+Dk^{2})^2}$. For the lower band $\left|-\right\rangle$, the Berry curvature $\Omega_{k\phi}^{-}$ is
\begin{equation}
\label{equation7}
\Omega_{k\phi}^{-}=\Omega_{\theta\phi}^{-} \frac{\partial(\theta,\phi)}{\partial(k,\phi)}	
= (\frac{\partial A_{\phi}^{-}}{\partial \theta}-\frac{\partial A_{\theta}^{-}}{\partial \phi}) \frac{\partial(\theta,\phi)}{\partial(k,\phi)}	=\frac{2 E C^{2} k^{3}}{\left[C^{2} k^{4}+\left(E+D k^{2}\right)^{2}\right]^{3 / 2}}.
\end{equation}
The Chern number of this spin-up band is
\begin{equation}
\label{equation8}
C_{\uparrow}=\frac{1}{2\pi}\int_{0}^{\infty}	\Omega_{k\phi}^{-} dk \int_{0}^{2\pi} d\phi = \text{sgn}(E)-\frac{D}{\sqrt{C^2+D^2}}.
\end{equation}
Therefore, for the continuous model of Eq. (\ref{equation1}), the spin Chern number $C_{S}$ is
\begin{equation}
\label{equation9}
C_{S}=\frac{1}{2}(C_{\uparrow}-C_{\downarrow})=	\text{sgn}(E)-\frac{D}{\sqrt{C^2+D^2}}.
\end{equation}

We note that continuous models involving the limit of infinite momentum may yield non-integer and thus non-physical spin Chern numbers. This can be resolved by introducing higher-order $k$-terms (such as a quadratic correction in the modified Dirac equation) or by mapping the continuous model onto a lattice tight binding model \cite{shen2011, kirtschig}. For the first option, with respecting $D_{6h}$ symmetry operations, we include the $k$-terms in Eq. (\ref{equation1}) up to quartic, $P (k_{x}^{2}+k_{y}^{2})^2 \sigma_{z}\otimes\tau_{z}$, and to sextic, $P (k_{x}^{2}+k_{y}^{2})^2 \sigma_{z}\otimes\tau_{z}+Q (k_{x}^{2}+k_{y}^{2})^3 \sigma_{z}\otimes\tau_{z}$. The derived spin Chern numbers are dependent on the highest order of the $k$-terms added, with $C_{S}=\text{sgn}(E)-\text{sgn} (P)$ for the quartic order and $C_{S}=\text{sgn}(E)-\text{sgn}(Q)$ for the sextic order. One may note that the sign of $E$ consistently contributes to the spin Chern number. 

Considering the second approach, i.e., mapping the continuous model to a tight binding model, a four-band Hamiltonian with the $p_{\pm}$ orbitals on a triangular lattice is constructed as
\begin{equation}
\label{equation10}
\begin{aligned}
H(\bm{k})=&\lambda_{0}\sigma_{z}\otimes\tau_{z}\\
&+2[\cos(k_{y})+2\cos(\frac{\sqrt{3}k_{x}}{2})\cos(\frac{k_{y}}{2})]\cdot[(t_{p\sigma1}+t_{p\pi1})\cdot\sigma_{0}\otimes \tau_{0} + \lambda_{1}\cdot \sigma_{z}\otimes\tau_{z}]\\
&+2(t_{p\sigma1}-t_{p\pi1})\cdot[-\cos(k_{y})+\cos(\frac{\sqrt{3}k_{x}}{2})\cos(\frac{k_{y}}{2})]\cdot\sigma_{0}\otimes\tau_{x} -   2\sqrt{3}(t_{p\sigma1}-t_{p\pi1})\cdot\sin(\frac{\sqrt{3}k_{x}}{2})\sin(\frac{k_{y}}{2}) \cdot\sigma_{0}\otimes\tau_{y} \\
&+2[\cos(\sqrt{3}k_{x})+2\cos(\frac{\sqrt{3}k_{x}}{2})\cos(\frac{3k_{y}}{2})]\cdot[(t_{p\sigma2}+t_{p\pi2})\cdot\sigma_{0}\otimes\tau_{0} + \lambda_{2}\cdot \sigma_{z}\otimes\tau_{z}]\\
&+2(t_{p\sigma2}-t_{p\pi2})\cdot[\cos(\sqrt{3}k_{x})-\cos(\frac{\sqrt{3}k_{x}}{2})\cos(\frac{3k_{y}}{2})]\cdot\sigma_{0}\otimes\tau_{x} -   2\sqrt{3}(t_{p\sigma2}-t_{p\pi2})\cdot\sin(\frac{\sqrt{3}k_{x}}{2})\sin(\frac{3k_{y}}{2}) \cdot\sigma_{0}\otimes\tau_{y}
\end{aligned}
\end{equation}
where the $\sigma$ and $\tau$ represent the Pauli matrices for spin and orbital degrees of freedom, respectively, as in Eq. (\ref{equation1}). $t_{p\sigma 1, p\pi 1}$ and $t_{p\sigma 2, p\pi 2}$ represent the first and second-nearest neighbor hopping parameters, respectively. $\lambda_{0}$ denotes the onsite SOC. $\lambda_{1}$ and $\lambda_{1}$ denote the first and second-nearest neighbor SOC, respectively. This lattice model is effectively mapped to the continuous model in Eq. (\ref{equation1}) as $A=6(t_{p\sigma1}+t_{p\pi1}+t_{p\sigma2}+t_{p\pi2})$, $B=\frac{3}{2}(t_{p\sigma1}+t_{p\pi1}+3t_{p\sigma2}+3t_{p\pi2})$, $C=\frac{3}{4}(-t_{p\sigma1}+t_{p\pi1}-3t_{p\sigma2}+3t_{p\pi2})$, $D=\frac{3}{2}(-\lambda_{1}-3\lambda_{2})$, and $E=\lambda_{0}+6\lambda_{1}+6\lambda_{2}$. The calculations of the $C_{S}$ for the valence bands in Eq. (\ref{equation10}) reference the code implemented in the PythTB package \cite{pythtb}. Integration of the Berry curvature is performed using a dense 101$\times$101 k-mesh. We test several sets of $\lambda_{0}$, $\lambda_{1}$, and $\lambda_{2}$, and find that as the sign of $E$ changes from positive to negative, the $C_{S}$ shifts from 0 to $-$2. As a result, the effective Hamiltonian in Eq. (\ref{equation1}) serves as a simplified and generalized model, capturing the essential relationship between the sign of $E$ and the topological phase transition.

\section{Band structures of RuF$_3$ and RuCl$_3$}
See Fig. \ref{fig:7} for the band structures of RuF$_3$ and RuCl$_3$.
\begin{figure}[htp]
  \centering
\includegraphics[width=17cm]{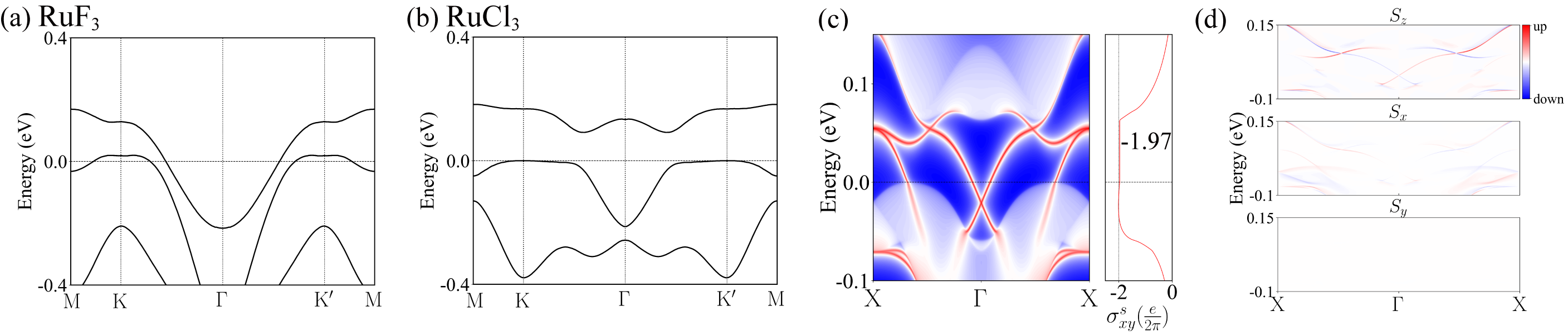}
  \caption{The band structures of monolayers (a) RuF$_3$ and (b) RuCl$_3$ in the nonmagnetic state. The monolayer RuF$_3$ exhibits metallic behavior, while RuCl$_3$, similar to RuBr$_3$ discussed in the main text, is an even spin Chern insulator. As shown in the (c) edge spectrum and SHC and (d) the spin components of edge states, RuCl$_3$ has two pairs of helical edge states, and its spin Hall conductance plateau within the bulk gap reaches a near-quantized value of $-$1.97.}
  \label{fig:7}
\end{figure} 

\end{widetext}

\bibliography{rsc} 

\end{document}